\documentclass{article}
\usepackage{dcase2022,amsmath,graphicx,url,times,booktabs, tabularx}
\usepackage{indentfirst}
\usepackage{amsfonts}
\usepackage{siunitx}
\usepackage{bm}
\usepackage[all]{nowidow}
\usepackage{multirow}
\usepackage[table]{xcolor}
\usepackage{diagbox}

\newcommand{\norm}[1]{\left\lVert#1\right\rVert}

\title{CoLoC: Conditioned Localizer and Classifier \\ for Sound Event Localization and Detection}

\name{S\l{}awomir Kapka\sthanks{Corresponding author.}, Jakub Tkaczuk}
\address{Samsung R\&D Institute Poland\\
      Warsaw, Poland \\
      s.kapka@samsung.com, j.tkaczuk@samsung.com}

\begin{document}

\ninept
\maketitle

\begin{sloppy}

\begin{abstract}
\indent \indent
In this article, we describe Conditioned Localizer and Classifier (CoLoC) which is a novel solution for Sound Event Localization and Detection (SELD). The solution constitutes of two stages: the localization is done first and is followed by classification conditioned by the output of the localizer. In order to resolve the problem of the unknown number of sources we incorporate the idea borrowed from Sequential Set Generation (SSG). Models from both stages are SELDnet-like CRNNs, but with single outputs. Conducted reasoning shows that such two single-output models are fit for SELD task. We show that our solution improves on the baseline system in most metrics on the STARSS22 Dataset.
\end{abstract}


\section{Introduction}
\label{sec:intro}

Sound Event Localization and Detection (SELD) is a complex task with many applications in robotics and surveillance. Since 2019, DCASE host annual Challenge in which Task3 is precisely SELD. This allows for gradual improvement of SELD systems over time. 

The original SELDNet \cite{adavanne2018sound} has a drawback of being unable to detect multiple overlapping occurrences of events from the same class. We follow \cite{cao2020event} and henceforth we call this problem \emph{homogeneous overlap}. Thus, a new track-wise output format has been introduced in \cite{cao2020event} incorporating Permutation Invariant Training (PIT) which precisely tackles the problem of homogeneous overlap. Since then, PIT was used in the improved version of DCASE2021 Top solution \cite{shimada2022multi} as well as in the DCASE2022 Task3 Baseline System \cite{dcase2022challengeweb}.

In this paper we propose a novel two-stage solution which incorporates class agnostic localizer based on Sequential Set Generation (SSG) and classifier conditioned on the output of the localizer. We will explain in the Section \ref{sec:method} how our solution solves the problem of homogeneous overlap without PIT. Our solution can be seen as an improved and refined version of our previous system from DCASE2019 \cite{kapka2019sound}. In this solution the estimator of the number of active sources is included as a part of a conditional localizer, i.e. using SSG localizer we can retrieve the number of active sound sources in each time frame. 

In the following Section \ref{sec:method} we describe the main components and the inference process of our method. In Section \ref{sec:training} we present a way how to train the components. Finally, in Section \ref{sec:evaluation} we describe detailed results obtained on the STARSS22 Dataset \cite{Politis2022starss22} and compare our solution with the baseline system.

\section{Proposed Method}
\label{sec:method}

\subsection{Motivation}
\label{ssec:motivation}

From the perspective of statistical learning theory, the optimal solutions to the problem of Empirical Risk Minimalisation (ERM) are conditional probability $P(Y|X)$ and conditional expectation $E(Y|X)$ for classification with cross-entropy loss and for regression with $L^2$ loss respectively \cite{vapnik1999nature}. By abuse of language, we will call the general solution to the ERM problem simply as conditional probability and we will use notations $P(Y|X)$ and $E(Y|X)$ rather frivolously.

SELD is the problem of simultaneous localization and detection, so the optimal solution in each time frame may be modeled as joint probability:
$$P(\{c_i\wedge l_i\}_{i=1..k}|X),$$
where $c_i$ and $l_i$ denote the class and the location of a detected event $e_i,$ and $k\leq N,$ where $N$ is the maximal number of overlapping events. The first problem rises from the fact that models do not output sets. The most popular workaround is to force a model to output a list of length $N$ denoting the individual tracks with class and location information, where some tracks may be empty.

In our solution we first localize all audio events using the Sequential Set Generation method, which simultaneously allows us to estimate the number of active sound sources. More precisely, the localizer returns, in a sequential manner, directions of arrivals (DOAs) conditioned by DOAs which it already returned starting from the empty set $\emptyset$:
\begin{align*} 
l_1 &=  E(l|X,\emptyset) \\ 
l_2 &=  E(l|X,\{l_1\}) \\
l_3 &=  E(l|X,\{l_1, l_2\}) \\
\dots\\
l_k &=  E(l|X,\{l_i\}_{i=1..k-1}) \\
\tau &=  E(l|X,\{l_i\}_{i=1..k}),
\end{align*}
where $\tau$ is a special token denoting that there is no more events. In our case $\tau$ is set to be an origin $\bm{0}$ from $\mathbb{R}^3.$

Based on the output from the localizer we then classify the event corresponding to this particular DOA. Thus informally
$$P(c_i\wedge l_i|X) = P(c_i|X,l_i)\cdot P(l_i|X),$$
which indicates that we could resolve the SELD task given an SSG localizer $E(l|X,\{l_i\}_i)$ and location-conditioned classifier $P(c|X,l).$ 

The only edge case where the above solution may fail is when two or more events overlap spatially. However, in practice outputs from models have temporal context which should resolve this issue.

\subsection{Stacked-Tracks}
\label{ssec:stackedtracks}

Let us consider a chunk of First Order Ambisonics (FOA) audio format in which there are at most $N$ overlapping audio events. With such audio we associate meta information about the location and classes of the occurrences in each time frame. By location we mean $xyz$ Cartesian coordinates on a unit sphere, and by class we mean one of the $K$ predefined classes. We aim to construct an $N\times T\times 4$ tensor \emph{stacked-tracks}, where $N$ is the maximal number of overlapping events, $T$ denotes the number of time bins and the last dimension contains information about locations and classes. The tensor contains all available meta information in a convenient form. To obtain stacked-tracks, we iterate sequentially over occurrences and stack them from bottom to top. If an event terminates in some track, then all events from the above tracks are stacked down. All remaining empty cells are filled with zeros in $xyz$ coordinates and with a new class index $K$ which is interpreted as the lack of any of predefined event, i.e. silence or unknown event. Figure 1 presents an example of how to obtain stacked-tracks. Since the tracks can be permuted before stacking, the stacked-tracks are not unique; we will exploit this during training later.

\begin{figure}[h]
\scriptsize
\centering
\begin{tabular}{|c||c|c|c|c|c|c|c|c|}
\hline
\multirow{2}{*}{Tracks} & \multicolumn{8}{c|}{Time Frames} \\
\cline{2-9}
 & 0 & 1 & 2 & 3 & 4 & 5 & 6 & 7\\
\hline
\multirow{4}{*}{T4} & & \cellcolor{blue!25}0.2 & \cellcolor{blue!25}0.2 & & & & & \\
& & \cellcolor{blue!25}0.7 & \cellcolor{blue!25}0.8 & & & & & \\
& & \cellcolor{blue!25}-0.2 & \cellcolor{blue!25}-0.1 & & & & & \\
& & \cellcolor{blue!25}3 & \cellcolor{blue!25}3 & & & & & \\
\hline
\multirow{4}{*}{T3} & & & \cellcolor{red!25}0.5 & \cellcolor{red!25}0.5 & \cellcolor{red!25}0.5 & \cellcolor{red!25}0.6 & \cellcolor{red!25}0.6 & \\
& & & \cellcolor{red!25}-0.7 & \cellcolor{red!25}-0.7 & \cellcolor{red!25}-0.7 & \cellcolor{red!25}-0.7 & \cellcolor{red!25}-0.7 & \\
& & & \cellcolor{red!25}0.5 & \cellcolor{red!25}0.5 & \cellcolor{red!25}0.5 & \cellcolor{red!25}0.4 & \cellcolor{red!25}0.4 & \\
& & & \cellcolor{red!25}7 & \cellcolor{red!25}7 & \cellcolor{red!25}7 & \cellcolor{red!25}7 & \cellcolor{red!25}7 & \\
\hline
\multirow{4}{*}{T2} & \cellcolor{green!25}-0.5 & \cellcolor{green!25}-0.4 & \cellcolor{green!25}-0.4 & \cellcolor{green!25}-0.4 & \cellcolor{green!25}-0.3 & & & \\
 & \cellcolor{green!25}0.6 & \cellcolor{green!25}0.7 & \cellcolor{green!25}0.7 & \cellcolor{green!25}0.8 & \cellcolor{green!25}0.8 & & & \\
 & \cellcolor{green!25}0.3 & \cellcolor{green!25}0.3 & \cellcolor{green!25}0.3 & \cellcolor{green!25}0.3 & \cellcolor{green!25}0.4 & & & \\
 & \cellcolor{green!25}3 & \cellcolor{green!25}3 & \cellcolor{green!25}3 & \cellcolor{green!25}3 & \cellcolor{green!25}3 & & & \\
\hline
\multirow{4}{*}{T1} & & & & & & & \cellcolor{yellow!50}0.7 & \cellcolor{yellow!50}0.7 \\
& & & & & & & \cellcolor{yellow!50}0.5 & \cellcolor{yellow!50}0.5 \\
& & & & & & & \cellcolor{yellow!50}-0.5 & \cellcolor{yellow!50}-0.5 \\
& & & & & & & \cellcolor{yellow!50}11 & \cellcolor{yellow!50}11 \\
\hline
\multirow{4}{*}{T0} & & & & \cellcolor{black!25}-0.9 &  \cellcolor{black!25}-0.9 &  \cellcolor{black!25}-0.8 & & \\
& & & &  \cellcolor{black!25}0.2 &  \cellcolor{black!25}0.2 &  \cellcolor{black!25}0.2 & & \\
& & & &  \cellcolor{black!25}0.1 &  \cellcolor{black!25}0.1 &  \cellcolor{black!25}0.2 & & \\
& & & &  \cellcolor{black!25}8 &  \cellcolor{black!25}8 &  \cellcolor{black!25}8 & & \\
\hline
\end{tabular}

\normalsize
\vspace{5pt}
$\downarrow\downarrow\downarrow \text{Stacking} \downarrow\downarrow\downarrow$
\vspace{5pt}

\scriptsize

\begin{tabular}{|c||c|c|c|c|c|c|c|c|}
\hline
\multirow{2}{*}{Tracks} & \multicolumn{8}{c|}{Time Frames} \\
\cline{2-9}
 & 0 & 1 & 2 & 3 & 4 & 5 & 6 & 7\\
\hline
\multirow{4}{*}{ST2} & 0.0 & 0.0 & \cellcolor{blue!25}0.2 & \cellcolor{red!25}0.5 & \cellcolor{red!25}0.5 & 0.0 & 0.0 & 0.0 \\
& 0.0 & 0.0 & \cellcolor{blue!25}0.8 & \cellcolor{red!25}-0.7 & \cellcolor{red!25}-0.7 & 0.0 & 0.0 & 0.0 \\
& 0.0 & 0.0 & \cellcolor{blue!25}-0.1 & \cellcolor{red!25}0.5 & \cellcolor{red!25}0.5 & 0.0 & 0.0 & 0.0 \\
& 13 & 13 & \cellcolor{blue!25}3 & \cellcolor{red!25}7 & \cellcolor{red!25}7 & 13 & 13 & 13 \\
\hline
\multirow{4}{*}{ST1} & 0.0 & \cellcolor{blue!25}0.2 & \cellcolor{red!25}0.5 & \cellcolor{green!25}-0.4 & \cellcolor{green!25}-0.3 & \cellcolor{red!25}0.6 & \cellcolor{red!25}0.6 & 0.0 \\
& 0.0 & \cellcolor{blue!25}0.7 & \cellcolor{red!25}-0.7 & \cellcolor{green!25}0.8 & \cellcolor{green!25}0.8 & \cellcolor{red!25}-0.7 & \cellcolor{red!25}-0.7 & 0.0 \\
& 0.0 & \cellcolor{blue!25}-0.2 & \cellcolor{red!25}0.5 & \cellcolor{green!25}0.3 & \cellcolor{green!25}0.4 & \cellcolor{red!25}0.4 & \cellcolor{red!25}0.4 & 0.0 \\
& 13 & \cellcolor{blue!25}3 & \cellcolor{red!25}7 & \cellcolor{green!25}3 & \cellcolor{green!25}3 & \cellcolor{red!25}7 & \cellcolor{red!25}7 & 13\\
\hline
\multirow{4}{*}{ST0} & \cellcolor{green!25}-0.5 & \cellcolor{green!25}-0.4 & \cellcolor{green!25}-0.4 & \cellcolor{black!25}-0.9 &  \cellcolor{black!25}-0.9 &  \cellcolor{black!25}-0.8 & \cellcolor{yellow!50}0.7 & \cellcolor{yellow!50}0.7 \\
& \cellcolor{green!25}0.6 & \cellcolor{green!25}0.7 & \cellcolor{green!25}0.7 &  \cellcolor{black!25}0.2 &  \cellcolor{black!25}0.2 &  \cellcolor{black!25}0.2 & \cellcolor{yellow!50}0.5 & \cellcolor{yellow!50}0.5 \\
& \cellcolor{green!25}0.3 & \cellcolor{green!25}0.3 & \cellcolor{green!25}0.3 &  \cellcolor{black!25}0.1 &  \cellcolor{black!25}0.1 &  \cellcolor{black!25}0.2 & \cellcolor{yellow!50}-0.5 & \cellcolor{yellow!50}-0.5 \\
& \cellcolor{green!25}3 & \cellcolor{green!25}3 & \cellcolor{green!25}3 &  \cellcolor{black!25}8 &  \cellcolor{black!25}8 &  \cellcolor{black!25}8 & \cellcolor{yellow!50}11 & \cellcolor{yellow!50}11 \\
\hline
\end{tabular}
\caption{An illustration of obtaining stacked-tracks from regular tracks. In this example there are five tracks with maximal overlap of three events. In this example there are 13 predefined classes, thus cells with coordinates at origin get a class label with index 13.}
\end{figure}

\subsection{Self-Conditioned SSG Localizer}
\label{ssec:selfcondlocssgloc}

The idea behind self-conditioned SSG localizer is to recursively localize all events from known classes without specifying class labels, i.e. it is class-agnostic. We will call this module simply as localizer. Our localizer consists of two trainable components:
\begin{itemize}
\item Localizer Encoder $L_{enc}:\mathbb{R}^3\to \mathbb{R}^c$, where $c$ is the hyperparameter denoting the number of new channels.
\item Localizer Network $L_{net}:\mathbb{R}^{t\times F\times C}\to\mathbb{R}^{T\times 3}$, where $C=c+c_f$ given that $c_f$ is the number of feature channels, where $t,F$ are numbers of time and frequency bins respectively, and where $T$ is the number of label time bins with meta resolution (in our case $t=5\cdot T$).
\end{itemize}

The core idea behind our self-conditioned SSG localizer is to recursively obtain all DOAs in all time frames. We start from the blank stacked-tracks and in each step we successively fill tracks in such a way that $L_{enc}$ encodes previously detected DOAs which $L_{net}$ should ignore. For that, we introduce below a modified version of the SSG method:

\textbf{Step 0.} During the first step we set $xyz$ to zeros in each time frame (is is convenient to look at it as putting empty row below ST0 in the Figure 1). Hence, we end up with the vector of size $t\times 3$. Then we apply $L_{enc}$ to it frame-wise to obtain tensor of size $T\times c$. Next, we repeat the tensor so many times to obtain vector which could be stacked with extracted features from audio channels. Thus, we obtain tensor of size $t\times F \times c$ which we concatenate with audio features tensor of size $t\times F \times c_{f}$ to obtain $t\times F \times C$ tensor. Finally, we feed it to $L_{net}$ to get the tensor of size $T\times 3$ denoting the predicted DOAs in each frame. When there is no active sound sources in a frame we expect it to return origin (i.e. $xyz=\bf{0}$) similarly as it was done in \cite{shimada2021accdoa}. On the other hand, when there are some events we expect the model to return $xyz$ coordinates of any of the occurring events. In the end we simply threshold length of vectors to decide whether there has been any event. If the length is greater than $0.5,$ then we conclude that there is an event, otherwise we put zeros. In summary, we obtain information about DOAs in each time-frame which we put in a first row ST0 in the blank stacked-tracks.

\textbf{Step 1.} During the second iteration, the output from the first step is fed into the $L_{enc}$. If in the first step in some frames an event was detected, then we encode these predicted DOAs, $\bf{0}$ otherwise. In the same way as in the first step, the encoded tensor is repeated and stacked with features channels. Now, we expect the $L_{net}$ model to output in each time-frame the DOA of second event if there is any and origin if there is none, or the already detected one is the only one. Analogously as in the first step, we threshold the length of output to decide if there is a new event. We stack the obtained results in stacked-track ST1 on top of ST0.

\textbf{Step n.} Lets say we already have $n-1$ stacked-tracks.. The aim of this step is to get the DOAs of events which haven't been already localized. For these frames where stacked-track no. $n-1$ denotes that there is less then $n-1$ events, we encode the origin via $L_{enc}$. Otherwise, for these frames where we acquired $n-1$ DOAs we encode them individually using $L_{enc}$ and for each such frame we average $n-1$ obtained embeddings from $\mathbb{R}^c$. Note that thanks to average-pooling (in contrast to max-pooling as it was resolved in \cite{zaheer2017deep}) we potentially preserve some additional information about DOAs count. So, we encoded the set of all previous DOAs which $L_{net}$ should ignore. Analogously as in previous steps, encoded embeddings are repeated and stack with extracted audio features. Next, we predict DOAs of new events if there are some, threshold lengths, and finally obtain a new stacked-track.

One may ask in which order the localizer should return DOAs. We did not impose any restrictions on that and let the model learn its own internal hidden order. It will be evident later from the training process how it is done. In essence, in each frame the localizer returns DOAs by DOAs in its own fashion until all sound sources have been localized.


\subsection{Location-Conditioned Classifier}
\label{ssec:loccondclass}

Self-Conditioned SSG Localizer outputs information about DOAs in each time frame written in the stacked-frame format. Location-Conditioned Classifier simply takes each row from stacked-tracks and in each time-frame outputs probabilities of predicted classes conditioned by DOAs. If the localizer predicted that there is no event in a frame, the classifier is conditioned by the origin and it is expected to predict additional special class with index $K$, where $K$ is the number of classes.

Our classifier similarly as localizer consists of two trainable components:

\begin{itemize}
\item Classifier Encoder $C_{enc}:\mathbb{R}^3\to \mathbb{R}^c$, where $c$ is the hyperparameter denoting the number of new channels (may be different than the one in localizer, but for simplicity we set it to be the same)
\item Classifier Network $C_{net}:\mathbb{R}^{t\times F\times C}\to\mathbb{R}^{T\times (K+1)}$, where notation is the same as in the localizer above and where $K$ denotes the number of classes. 
\end{itemize}

\section{Training process}
\label{sec:training}

In our solution localizer and classifier are trained completely separately.

\subsection{Self-Conditioned SSG Localizer}
\label{ssec:selfcondloctrain}

Let's say we have an audio chunk and associated meta with $N$ stacked-tracks, where $N$ denotes the maximal number of overlapping events. We select a random integer $r$ from $0$ to $N-2$ in a uniform way and split stacked-tracks into two parts:
\begin{itemize}
\item Conditioning part containing tracks from $0$ to $r-1$,
\item Target part containing tracks from $r$ to $N-1$.
\end{itemize}
The aim of this splitting is to imitate the $r$'th iteration from the inference, by hiding stacked-tracks with indices $\geq r$. 


We feed the Conditioning part into $L_{enc}$ frame-wise in the following way: if there is some event in $r-1$'th stacked-track, then we encode all DOAs via $L_{enc}$ and avarage pool the embeddings. Otherwise, we encode $\bm{0}$ via $L_{enc.}$ We then repeat obtained embeddings so many times to be able to stack them with audio features and we feed an acquired tensor through $L_{net}$. Thus, in each time-frame $s$ we obtain new $xyz$ coordinates which we denote by $l^s_{pred}$. We compare the predicted coordinates with the ones from the Target part. In each time frame we compute a $L_{1.5}$ distance between predicted DOAs and the ground truth DOAs from Target part and set the minimum value as our loss. If there are no more active sound sources in Target part we enforce the target to be a zero vector by minimizing $L_{1.5}$ norm of the predicted DOA. I.e.
\scriptsize
$$Loss_{loc}^{r, k}(\bm{l}^s_{gt}, l^s_{pred}) = 
\begin{cases}
  \min_{i=r..k}\norm{l^s_{{gt}_i} - l^s_{pred}}_{1.5} & \text{if $k\geq r$} \\
  \norm{l^s_{pred}}_{1.5} & \text{if $k < r$}
\end{cases},$$
\normalsize
where $k$ is the index of the last nonzero DOAs from the Target part of stacked-tracks. During training, given a batch of size $B$ of audio-meta pairs we select random $r$ for each item in the batch, compute $Loss^{r, k}_{loc}$ loss (frame-wise and item-wise) and average it over all $k\leq N$ and $r\leq k$. I.e. the final loss between $[\bm{l}_{gt}]_b^t$ and $[l_{pred}]_b^t$ is
\scriptsize
$$\frac{2}{N(N+1)+2}\sum_{k\leq N}\sum_{r\leq k}\frac{1}{|T_{r,k}|}\sum_{s\in T_{r,k}}Loss_{loc}^{r,k}(\bm{l}^s_{gt}, l^s_{pred}),$$
\normalsize
where $T_{r,k}$ stand for time-frames among whole batch and where $k,r$ were sampled according to the rule described above.

One may ask why we decided to select $L_p$ norm with $p=1.5$. Since for $p=2$ the optimal solution is the expected value, there is a possible risk of ignoring the conditioning $L_{enc}$ and averaging $xyz$ outputs in the case of multiple overlapping sound sources. Conversely, for $p=1$ the optimal solution is the median which may be too "sharp" decision making. Thus $p=1.5$.

\subsection{Location-Conditioned Classifier}
\label{ssec:selfcondtrain}

In the same manner as in the case of the localizer, let's assume that we have some audio-meta pair from a chunk of the sound signal and let $N$ denote the maximal number of overlapping events. We select random number $r$ from $0$ to $N-1$ and we select $r$'th row from stacked-meta containing DOA information. We aim to output classes associated with these DOAs. In each time-frame we encode the DOAs from the selected row via $C_{enc}$. Then we repeat the encoded tensor and stack it with audio features. We forward it through $C_{net}$ to obtain $K+1$ scores in each time-frame $s$ associated with $K$ classes and one score denoting unknown class or lack of any event. We then simply compare predicted probability scores $p^s_{pred}$ with ground truth classes $c^s_{gt}$ from the selected stacked-track. Since there is great imbalance of classes due to the fact that class associated with unknown event is over-represented, we used focal loss \cite{lin2017focal} with $\gamma=1$ instead of regular cross-entropy. I.e.
\scriptsize
$$Loss_{cls}(c^s_{gt}, p^s_{pred}) = - (1-p^s_{c,pred})\log(p^s_{c,pred}).$$
\normalsize
For a batch of size $B$ and $T$ time-frames per item we simply average everything to obtain the final loss for a back-propagation.

\subsection{Angle perturbation}
\label{ssec:angleperturbation}

One drawback of our approach is the error propagation. Since localizer and classifier are conditioned by the output of the localizer, the error in DOA predictions downgrade the quality of outputs in the next steps. To partly resolve that issue we decided to perturb angles fed to $L_{enc}$ and $C_{enc}$ during the training as well during the inference. We decided to randomly perturb azimuth and elevation by 5 degrees in each time-frame from meta.

\subsection{Stack-track permutation}
\label{ssec:trackpermutation}

In order to increase the number of training samples we could permute ordinary tracks before stacking into stacked-tracks. However, in our case it was more convenient to operate on stacked-tracks. Thus, for each stacked-track tensor we permute the tracks and stack them back.

In our solution we constrained to $N=3$, i.e. we allow up to 3 overlapping audio events. We simply ignore the rest.

\subsection{Model Architecture}
\label{ssec:modelarchitecture}

For $L_{net}$ and $C_{net}$ we used the SELDnet-like architecture introduced in \cite{adavanne2018sound}. More precisely, for localizer and classifier we took the baseline model \cite{Politis2022starss22} and changed the number of filters in convolutional blocks from 64 to 128. After the last time-distributed dense layer we put at the end another time-distributed dense layer with $3$ outputs with tanh activation for localizer and with $13$ outputs with softmax activation for classifier.

We set $L_{enc}$ and $C_{enc}$ to be single dense layers with $3$ inputs and $c=5$ outputs with tanh activation. 

As for the models complexity, $L_{net}$ and $C_{net}$ have slightly above $2.3$ million parameters each while $L_{enc}$ and $C_{enc}$ have just $20$ parameters each.

\subsection{Complexity}
\label{ssec:complexity}

During the training of the localizer, we need to compute up to $\frac{N\cdot(N+1)}{2}+1$ components to the loss given that the maximal number of overlapping sound sources is $N.$ For the classifier, the complexity scales linearly with $N.$ This contrasts with PIT, where in principle we need to compute $N!$ components to the loss. During the inference as described in Section \ref{sec:method}, our method requires up to $N$ steps for both localizer and classifier.

\subsection{Features and augmentations}
\label{ssec:features}

We used 24kHz FOA format for our nets. We extracted complex spectrograms from each four FOA channels using Short Time Fourier Transform (STFT) with $n_{fft}=1024$ and Hanning window and hop length of $960$ and $480$ respectively. From each obtained spectrogram we acquire the log-power spectrogram and the phase spectrogram. For the last three channels we used intensity vectors as it was done in \cite{Cao2019}. In summary, from each FOA audio signal we acquire $11$ audio features of size $t\times 513,$ where $t$ is the number of time bins from the STFT. In our case we randomly selected 5s audio chunks from the recordings which constitutes of $250$ time bins. 

For data augmentation we used volume perturbation by selecting a random number between $0.5$ and $1.5$ and multiplying all audio channels by that number. We also used FOA domain spatial augmentation \cite{mazzon2019sound} to augment every fourth audio-meta pair. 

We trained localizer and classifier using Adam optimizer \cite{Adam} with default parameters except learning rate which we set to be $0.0005$. We trained both models for half a million batches of size 96.

\section{Evaluation}
\label{sec:evaluation}

\subsection{Metrics}
\label{ssec:metrics}

The DCASE2022 Challenge Task3 organizers provided two types of datasets for the development stage \cite{Politis2022starss22}:
\begin{itemize}
\item Synth: 1200 one-minute synthesized mixtures from collected SRIRs and selected sound events from FSD50K \cite{fonseca2021fsd50k} 
\item STARSS22: 292 minutes of real recordings simulating real life scenarios gathered in 11 rooms in Tokyo and Tempere. 
\end{itemize}
STARSS22 is further split into train and test folds. In this section we will discuss results on the test fold. During the training, for every batch we sampled half of the recordings from Synth dataset and half from the train split from STARSS22, utilizing all the data. We noticed that if we do not use synthetic dataset the scores drop drastically.

For localizer we first report the average DOA error in angles. In Table \ref{tab:doaerror} we show average DOA errors in multiple cases differentiating between number of active sound sources and number of DOAs in conditioning. As excepted, the more DOAs in conditioning, the larger the error. However, what is interesting is the fact that the more sound sources, the more accurate the model is in detecting the first few DOAs.

\begin{table}[h]
\caption{Dependence of the localizer's average DOA error (in degrees) on the number of active sources (noas) and the number of DOAs in conditioning (\#cond) on STARSS22 test split.} 
\centering
\begin{tabular}{c|ccc}
\toprule
\diagbox{noas}{\#cond} & 0 & 1 & 2\\
\hline
     1 & 23 & & \\
     2 & 20 & 33 & \\
     3 & 14 & 19 & 51 \\
\bottomrule

\end{tabular}
\label{tab:doaerror}
\end{table}
\normalsize

For the classifier we report conditional accuracy (CAcc) in which we count the class $c$ with maximal probability $P(c|X,l)$ correct if that class corresponds to the conditioning DOA $l$. Otherwise we treat it as an incorrect prediction. On the STARSS22 test split we achieved CAcc of 68\%. We also kept track of the number of frames where the classifier misses known classes and of frames where the classifier predicted a known class where there is none, but for those cases the classifier achieve almost perfect scores.

Finally, we report our results on official DCASE2022 Task3 metrics \cite{Politis2022starss22}, namely: the localization-dependent error rate $ER_{20^\circ}$, F1-score $F_{20^\circ}$, the localization error $LE_{CD}$ and the localization recall $LR_{CD}.$ We compare the baseline system with two versions of our solution. In the first one we will predict up to 3 DOAs and corresponding classes, and in the second one we terminate inference on 2 events. We denote these solutions as $\verb|max_ov3|$ and $\verb|max_ov2|$ respectively. Note that we used the very same models in both solutions, only the inference changes. We summarise the results in Table \ref{tab:scores}. In most metrics our solution outperforms the baseline system. The only one when our solution is lacking is the error rate. Furthermore, the error rate is worse when we try to infer more events. We speculate that this is due to the fact that the localizer is very inaccurate in later steps (see Table \ref{tab:doaerror}), which may generate many false positives.

\begin{table}[h]
\caption{Official metrics; the \textbf{boldface} denotes the best scores.}
\centering
\begin{tabular}{l||cccc}

\toprule
 & $ER_{20^\circ}$ & $F_{20^\circ}$ & $LE_{CD}$ & $LR_{CD}$\\
\midrule
$\verb|Baseline|$ & $\bm{0.71}$ & $21\%$ & $29.3^\circ$ & $46\%$ \\
$\verb|max_ov3|$ & $0.85$ & $32\%$ & $24.7^\circ$ & $\bm{51\%}$ \\
$\verb|max_ov2|$ & $0.76$ & $\bm{33\%}$ & $\bm{24.6^\circ}$ & $49\%$ \\
\bottomrule

\end{tabular}
\label{tab:scores}
\end{table}
\normalsize

\section{Conclusion}
\label{sec:conclusion}

In this paper we presented alternative solution to the SELD problem. Our solution uses our custom SSG method to determine DOAs one by one and as a result to determine number of sound sources in each time frame. The localizer is followed by the location-conditioned classifier. The performance of our method is comparible to the DCASE2022 Task3 baseline system which uses PIT. Thus, we imagine that in the future SSG and PIT may be combined together to obtain the best of both worlds.





\pagebreak

\bibliographystyle{IEEEtran}
\bibliography{refs}
\end{sloppy}
\end{document}